# PANDA-film: an automated system for electrodeposition of polymer thin films and their wetting analysis

Harley Quinn,[a] Gregory A. Robben,[a] Zhaoyi Zheng,[a] Jin Yan,[a] Yuanzhi Li,[b] Zhaoji Yang,[a] Maria Politi,[c] Nadya Peek,[d] Lilo Pozzo,[e] Jörg G. Werner,[*abf] and Keith A. Brown[*abg]

[a]Division of Materials Science & Engineering, Boston University, Boston, MA, 02215 USA.

[b]Department of Mechanical Engineering, Boston University, Boston, MA, 02215 USA

[c]Department of Chemistry, University of British Columbia, Vancouver, BC V6T 1Z1 Canada

[d]Department of Human Centered Design and Engineering, University of Washington, Seattle, WA 98195 USA

[e]Department of Chemical Engineering, University of Washington, Seattle, WA 98195 USA

[f]Department of Chemistry, Boston University, Boston, MA, 02215 USA

[g]Department of Physics, Boston University, Boston, MA, 02215 USA

*E-mail: brownka@bu.edu, jgwerner@bu.edu

**Abstract**

Thin polymer films are widely used as functional and protective coatings. However, determining the composition and processing conditions that produce a desired function is a tedious process due to the large number of factors that must be considered and the manual nature of most synthesis and characterization methods. Self-driving labs (SDLs), or robotic systems that prepare and test materials samples, are designed to overcome this bottleneck by enabling the efficient exploration of complex parameter spaces. In this paper, we report the development and testing of the polymer analysis and discovery array (PANDA)-film, a modular SDL for electrochemically synthesizing polymer films and then determining their water contact angle as a measure of surface energy. The system is designed to be highly modular and based upon a low-cost gantry platform to facilitate adoption. In addition to validating fluid handling and electrochemical tasks, we introduce two novel modular capabilities that enable PANDA-film to run sustained campaigns to study the wetting properties of films: (1) an electromagnetic capping/decapping system to mitigate fluid evaporation, and (2) a top-down optical method to determine water contact angle based upon reflectance. These capabilities are validated by depositing and characterizing a poly(allyl methacrylate) (PAMA) film using electrodeposition of polymer networks (EPoN). Comprehensive details for replicating the hardware and software of PANDA-film are included.

**1. Introduction**

Autonomous research platforms have gained traction in materials science as a way to streamline experimentation and accelerate discovery of new materials.[1,2] By combining automated synthesis and characterization with algorithmic decision-making, these systems, referred to as self-driving labs (SDLs), can iteratively explore large parameter spaces in a closed loop with minimal human intervention. In addition to accelerating optimization, SDLs offer benefits such as data quality, reproducibility, and standardization of experimental protocols. Recent examples of autonomous experimentation systems span diverse areas including additive manufacturing,[3] adhesive materials,[4] batteries,[5] catalysis,[6] electronic polymers,[7] photovoltaics,[8] and thin films,[9] each tailored to address the specific challenges within its domain. In parallel, platforms such as Science Jubilee[10] have demonstrated how open-source and modular designs can enable broad adoption and flexibility across many materials domains. However, realizing their full potential for a given application requires SDLs with application-specific hardware and workflows designed explicitly for measuring the properties of interest.[11]

One emerging, yet underexplored, direction is the use of SDLs to investigate surface phenomena, particularly in films where surface chemistry dominates interfacial properties and can dictate performance. Polymer films prepared using electrodeposition are especially attractive due to their ability to form thin, conformal coatings with tunable properties.[12–14] Autonomous systems such as PANDA and Polybot have successfully studied polymer film deposition for electronic applications by focusing on properties like electrochromic switching behavior,[15] or conductivity and uniformity,[7] while similar efforts have also extended to emulsion polymerization.[16] Characterization of thin films is a persistent challenge as bespoke hardware must often be developed to measure the property of interest. Automated image-based methods, including deep-learning models like DeepThin,[17] have shown the potential of image-based techniques to detect defects such as cracking or dewetting in polymer thin films from optical images, illustrating the feasibility of automated surface analysis.[9,18] However, these methods primarily focus on morphology and capturing chemical surface features, such as functionalization or wetting behavior, remains significantly more challenging to implement in a high-throughput, automated context.

Contact angle measurements offer a simple and powerful route to infer surface energy and liquid spreading behavior, which is of importance for surface functionalization or multilayer coatings. Protocols based on the Young–Dupre and Fowkes models are commonly used to determine surface energy for a variety of polymer substrates, including polycarbonate and polyethylene[19,20] and contact angle analysis is a standard technique for evaluating surface wettability and surface chemistry.[21,22] Because wetting behavior can also reflect the density and type of reactive surface sites, it could potentially offer a practical screening tool for chemical modification. Extending this rationale to electrodeposited films, surface properties can be tailored for specific applications in sensing,[13] coatings,[23,24] and biointerfaces[25] where precise control over surface chemistry is essential. Dynamic changes in contact-angle measurements including how contact angles evolve when the liquid-solid-vapor interface is in motion (e.g., during droplet spreading or dewetting) and how angles change with time following surface modifications (e.g., treatments, aging, adsorption) have also been shown to sensitively capture surface modification or chemical treatment effects by detecting changes in surface roughness, chemical heterogeneity, or contamination.[26,27] Despite their utility, contact angle measurements are rarely implemented in automated workflows. This stems from the fact that most contact angle goniometers rely on a side-view geometry, in which a droplet profile must be imaged and fit to extract the contact angle. While accurate, this approach is not readily compatible with high-density sample formats or robotic handling because each droplet must be oriented toward a camera at a precise angle and often requires manual focusing, which renders array-type sample layouts incompatible.

By contrast, a top-down approach circumvents these limitations. Imaging droplets from above allows the use of standard well-plate layouts, where many samples can be addressed in parallel, and eliminates the need for complex side-view optics or individual droplet alignment. This geometry is also more amenable to robotic dispensing, automated motion platforms, and surface mapping across large arrays. In addition, the image data are straightforward to process with automated scripts, enabling integration into closed-loop or high-throughput workflows. At the same time, robust top-down measurements, especially for hydrophobic regimes (contact angles >90°) require precise control of optical geometry,[28,29] drop shape reconstruction,[30,31] and calibration,[28,29] particularly on rough or curved surfaces.[32,33] While top-down methods unlock well-plate compatibility and automation, they have yet to be fully incorporated into autonomous experimentation platforms.

To address these challenges, we introduce PANDA-film, a system specifically configured for investigating thin-film surface properties. PANDA-film is a reconfiguration and advancement of the modular PANDA platform[15] and uses a custom well-plate format with a transparent conductive bottom that enables optical access from above and below the well plate for high-resolution droplet imaging and allows for precise electrochemical control. The system features transformative innovations beyond the original PANDA system including an automated vial capping/decapping system to mitigate evaporation of stock reagents, a common barrier in liquid handling for autonomous platforms, and a novel top-down method to measure water contact angle. We validate the platform's performance across key tasks including fluid dispensing and optical imaging. We then demonstrate its ability to detect subtle differences in surface chemistry through measurements of their water contact angles. To explore the system's utility, we electrodeposit films of poly(allyl methacrylate) (PAMA) with randomly distributed phenolic side groups[34] in replicates under well-controlled conditions (i.e., using the same stock polymerization solution, deposition potential, deposition time, and electrode composition) and perform subsequent water contact angle measurements.

## 2. Experimental Methods

### 2.1 Chemicals and materials

For experiments to study fluid handling, de-ionized water (18.2 MΩ·cm Milli-Q, Millipore) was used. For the initial contact angle calibration, glass substrates (25 × 25 mm$^2$) were prepared with the following surface coatings: polymethyl methacrylate (PMMA; 1000 HARP eB 0.3, KemLab) via spin coating; Microposit S1813 G2 positive photoresist (electronic grade, Rohm and Haas) via spin coating; polydimethylsiloxane (PDMS; Sylgard 184, Electron Microscopy Sciences) in a 10:1 base-to-crosslinker ratio via doctor blading; and SU-8 (Formulation 2, Kayaku Advanced Materials) via spin coating. In addition, indium tin oxide (ITO) (EJUITOX403A4, Kurt J. Lesker Company) and tungsten oxide (WO$_3$) (EJUWOXX303A4, Kurt J. Lesker Company) films were deposited via DC sputtering (EVOVAC, Angstrom Engineering). Bare glass and freshly cleaned glass substrates (rinsed with isopropanol and dried under a nitrogen stream) were also included.

**Polymer synthesis**

A random copolymer of allyl methacrylate (AMA) with a small fraction of glycidyl methacrylate (GMA) was synthesized via living anionic polymerization. Briefly, for a typical synthetic procedure, the monomers (both from Sigma-Aldrich) were purified with calcium hydride (Sigma-Aldrich), freeze-pump-thawed three times, and distilled using a high-vacuum Schlenk line prior to use. For the polymerization of poly(allyl methacrylate-*r*-glycidyl methacrylate) (P(AMA-*r*-GMA)) in tetrahydrofuran (THF, anhydrous, 99.8%, Sigma-Aldrich, purified with n-butyl lithium/1,1-diphenyl ethylene and distilled), 1,1-diphenyl ethylene (97%, Sigma-Aldrich, purified with n-butyl lithium and distilled) was first added followed by the initiator sec-butyllithium (solution in cyclohexane, Sigma-Aldrich). The resulting red 1,1-diphenyl hexyl lithium solution was cooled to -78 °C with a dry-ice bath of acetone and a predetermined ratio of AMA and GMA monomer mixture (19:1 by mol) was subsequently added. The solution was vigorously stirred for 2.5 h until degassed methanol (≥99.9%, HPLC, Fisher Chemical) was added to terminate the polymerization. Subsequently, the solvent was removed from the mixture using a rotary evaporator and the resulting solid was redissolved in chloroform (anhydrous, ≥99%, Sigma-Aldrich) at approximately 20 wt% followed by two cycles of precipitation into cold methanol, filtration, and redissolution. The purified product was dried in a vacuum oven at 50 °C for 3 days to yield P(AMA-*r*-GMA) as a white powder. The copolymerization was confirmed using proton-nuclear magnetic resonance ($^1$H-NMR) spectroscopy in deuterated chloroform on an Agilent 500 MHz VNMRS spectrometer, and the AMA:GMA molar ratio was quantified at 94:6 using the integrated areas of the respective side-group protons (Figure S1 bottom).

**Phenol modification of PAMA-co-PGMA**

2.6 g of PAMA-*r*-PGMA was dissolved in 16 mL of dimethyl sulfoxide (DMSO, anhydrous, ≥99.9%, Fisher Chemical). Triethylamine (TEA, 212.5 μL, Sigma-Aldrich) was added beneath the solution surface under stirring. Separately, 4-mercaptophenol (192.3 mg, 97%, Sigma-Aldrich) was dissolved in 4.0 mL DMSO and added dropwise to the polymer solution using a syringe. The reaction mixture was degassed with argon for 3 min and stirred at 70 °C in an oil bath. After 18 h, 87.3 μL of acetic acid (Sigma-Aldrich) was added to quench the reaction (1:1 molar ratio to TEA). Subsequently, the crude reaction mixture was precipitated into a 1 M aqueous sodium chloride solution, centrifuged, and washed with methanol (≥99.9%, HPLC, Fisher Chemical). The resulting phenol-modified polymer was dried in a vacuum desiccator for 3 days to yield a colorless solid. The successful modification (100% addition of 4-mercaptophenol to the GMA comonomer) was confirmed using $^1$H-NMR spectroscopy in deuterated chloroform on an Agilent 500 MHz VNMRS spectrometer (Figure S1 top). The resulting polymer can be considered PAMA modified with 6% phenolic comonomers ("phenol-modified PAMA").

**Phenol-modified PAMA stock solution**

The stock solution of phenol-modified PAMA (200 mg/mL) was prepared by dissolving 2.0 g of the polymer in 10 mL of 0.5 M tetra-n-butylammonium perchlorate (TBAP, Thermo Scientific) in N,N-dimethylformamide (DMF, ≥99.8%, Fisher Chemical). Triethanolamine (TEAA, Sigma-Aldrich) was added as a base at a 1:1 molar ratio relative to the phenol groups.

**2.2 Working electrode fabrication and electrochemical measurements**

All polymer films were electrodeposited on custom transparent conductive 40-well plates, which were fabricated by bonding molded polydimethylsiloxane (PDMS) gaskets to glass slides (86 × 126 mm²) that had been coated with ITO using DC sputtering. The gaskets were attached using a thin layer of uncured PDMS and cured for 48 hours at room temperature. While prior work utilized laser-cut PDMS gaskets,[15] this study employed molded PDMS gaskets to streamline fabrication. To this end, a custom 3D-printed mold was used to define the wells, eliminating several post-processing steps.

Electrochemical experiments were performed using a PalmSens EmStat4S potentiostat in a three-electrode configuration. The ITO-coated glass bottoms of the wells functioned as the working electrode, a platinum wire (0.25 mm diameter, 99.9% trace metals basis) served as the counter electrode, and an Ag wire (Sigma Aldrich) was used as a pseudo-reference electrode (AgQRE). Polymer films were deposited from the stock solution using a constant potential of 1.8 V vs. AgQRE for 600 s with chronoamperometry. After deposition, films were sequentially rinsed with DMF, rinsed with acetonitrile (anhydrous, Fisher Scientific), and then allowed to dry prior to contact angle measurements.

**2.3 Film thickness measurements**

Film thickness was determined using a Filmetrics F20 reflectance system (Filmetrics Inc.) with a spectral range of 380–1050 nm. Measurements were performed in a normal-incidence geometry. A multi-layer optical model was used, consisting of a transparent polymer film deposited on an ITO-coated glass substrate. The polymer layer was fit using a Cauchy dispersion model, while the optical constants of the ITO and glass substrate were fixed to known values. Thicknesses are reported as the mean of five locations per sample.

**2.4 PANDA-film System Architecture**

PANDA-film is a configuration of the modular PANDA platform, designed for electrodeposition of polymer networks (EPoN)[12,34,35] and surface characterization. The system is built on a modified CNC gantry (PROVerXL 4030 V2) with a 300 × 400 mm² working area, which provides a rigid and repeatable framework for automated positioning. The stock spindle was replaced with a custom 3D-printed modular tool holder, referred to as the process automation widget (PAW) (Figure 1A). The PAW establishes a standardized interface for mounting and swapping components, allowing PANDA-film to integrate different tools without permanent modification to the gantry. The PAW houses six primary modules (Figure 1B): (1) Electrochemical cell – a platinum wire counter electrode coiled around a glass capillary housing an Ag wire pseudo-reference electrode that, when paired with the working electrode mounted on the deck, enable controlled three-electrode electrochemistry and electrodeposition within each well. (2) Electromagnetic decapper (5V 50N, uxcell) – with an integrated cap detection sensor (2168, Adafruit) to automate vial sealing and unsealing, reducing solvent evaporation from stock vials. (3) Imaging system – a telecentric lens (#52-271, Edmund Optics) coupled to a C-mount camera (Grasshopper 3, FLIR), providing distortion-free droplet and film imaging. (4) Robotic pipette – a single-channel OT-2 P300 pipette (Opentrons) controlled via an Arduino for precise liquid handling. (5) Lighting components – a NeoPixel light ring (2863, Adafruit) for uniform illumination imaging and programmable button LEDs (NeoPixel 4776, Adafruit) positioned to create the reflection geometry for contact-angle analysis. (6) Embedded control – an Arduino Uno SMD R3 for direct control of PAW components (pipette actuation, decapper, LEDs) and communication with the higher-level orchestration software.

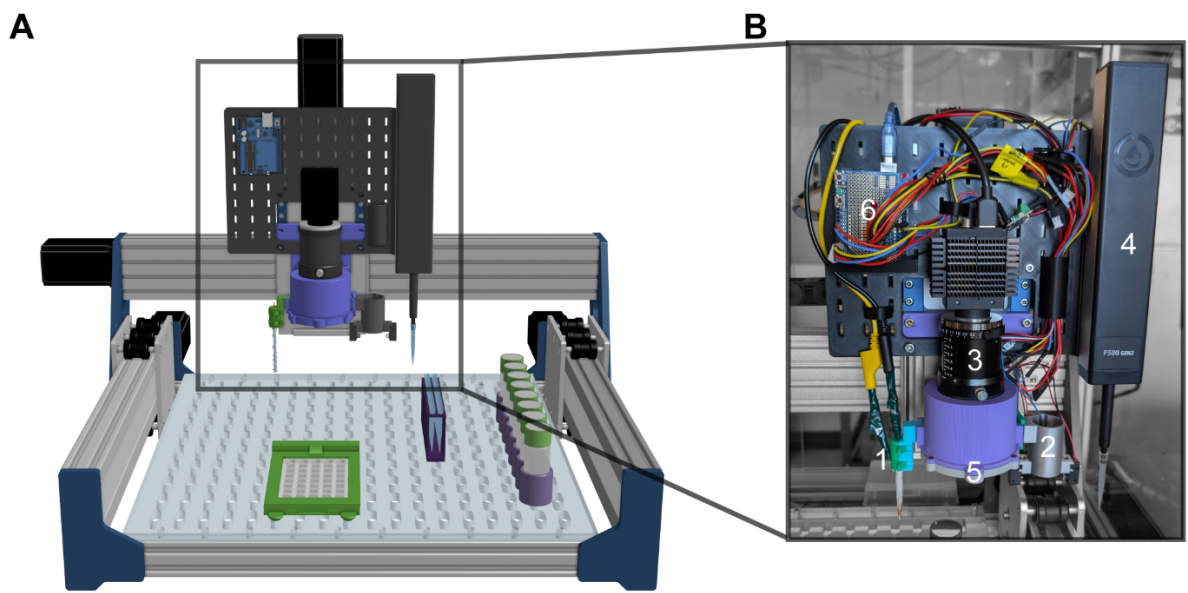

**Figure 1: A. CAD drawing of the polymer analysis and discovery array (PANDA)-film that highlights the process automation widget (PAW) mounted to the CNC gantry. B. Photograph of the six modules on the PAW: (1) reference and counter electrodes, (2) electromagnetic capper/decapper along with IR break beam sensor, (3) imaging system including a telecentric lens and camera, (4) robotic pipette, (5) lights including a ring light and button LEDs, (6) Arduino embedded control system.**

Accessories including stock solution holders, rinse vials, waste containers, a pipette tip rack, and a custom substrate mount were secured to the gantry deck using custom pill-slot mounting points (Figure S2). These mounting points allow rapid reconfiguration of the deck for different experiments while maintaining consistent and predictable accessory placement; positions can be easily calculated from their slot coordinates. A process-monitoring camera was mounted above the system to provide real-time visualization and video recording of experimental progress (Figure S3). For pipette validation experiments, the deck was temporarily removed to accommodate an analytical balance (Entris II Essential Precision Balance, Sartorius) for an estimate of volume through mass measurement. Pipetting functions, including aspiration, dispensing, and tip ejection, were controlled by an Arduino mounted on the PAW in combination with a TMC2209 stepper motor driver (6121, Adafruit), while spatial positioning of the pipette was achieved using the CNC gantry. Electrochemical experiments were performed using a PalmSens EmStat4S potentiostat, which interfaced directly with the control software. Full circuitry diagrams for the Arduino and pipette control are provided in the Supplementary Information (Figures S4 and S5).

## 2.5 Software

PANDA-film operates using the same text-based terminal interface described in previous work,[15] providing a flexible method for executing experimental protocols and monitoring system status. PANDA-film is hosted on a Raspberry Pi and accessed remotely via SSH from a centralized PC, enabling headless operation of the instrument, reducing reliance on a dedicated local machine, and allowing multiple systems to be coordinated from a single workstation. The software stack integrates hardware drivers for the pipette, potentiostat, imaging system, and auxiliary components through a modular set of Python libraries, making it straightforward to extend functionality or reconfigure hardware without modifying core code. Logging and metadata capture are handled automatically during each experiment using an SQL database interfaced with drivers from the panda-lib Python library. This ensures data integrity, as well as reproducibility of experiments and facilitates downstream analysis. Additional details on the software architecture, along with all source code used in this work, are provided in a repository as described as described in the data availability statement.

## 3. Results and Discussion

### 3.1 PANDA-film System Performance and Reproducibility

To build confidence in the development of the PANDA-film, we systematically evaluated each newly integrated module with a focus on reproducibility and robustness. One noted limitation of the original PANDA was that stock vials were open to the air, leading to evaporation, shortened campaign lengths, and necessitating manual replenishment. Thus, a primary objective for PANDA-film was to minimize evaporation and extend unattended operation time. We addressed this by incorporating a capping/decapping module. While commercial capping/decapping systems rely on complex grippers capable of grasping and rotating caps, we hypothesized that a simpler approach, using the magnetic pickup of custom-designed caps without rotary motion, would be easier to implement. The module consists of an electromagnet mounted in a 3D-printed holder in conjunction with an infrared (IR) break beam sensor to detect successful cap pickup and replacement (Figure 2A). Custom-designed caps were fabricated by 3D printing polylactic acid (PLA) using a Bambu A1 printer, filling the interior with PDMS to provide chemical resistance, and affixing a steel disc with 3M adhesive to a recessed region on the top of each cap. A cross-sectional view of the final caps and capper/decapper assembly is shown in Figure 2B.

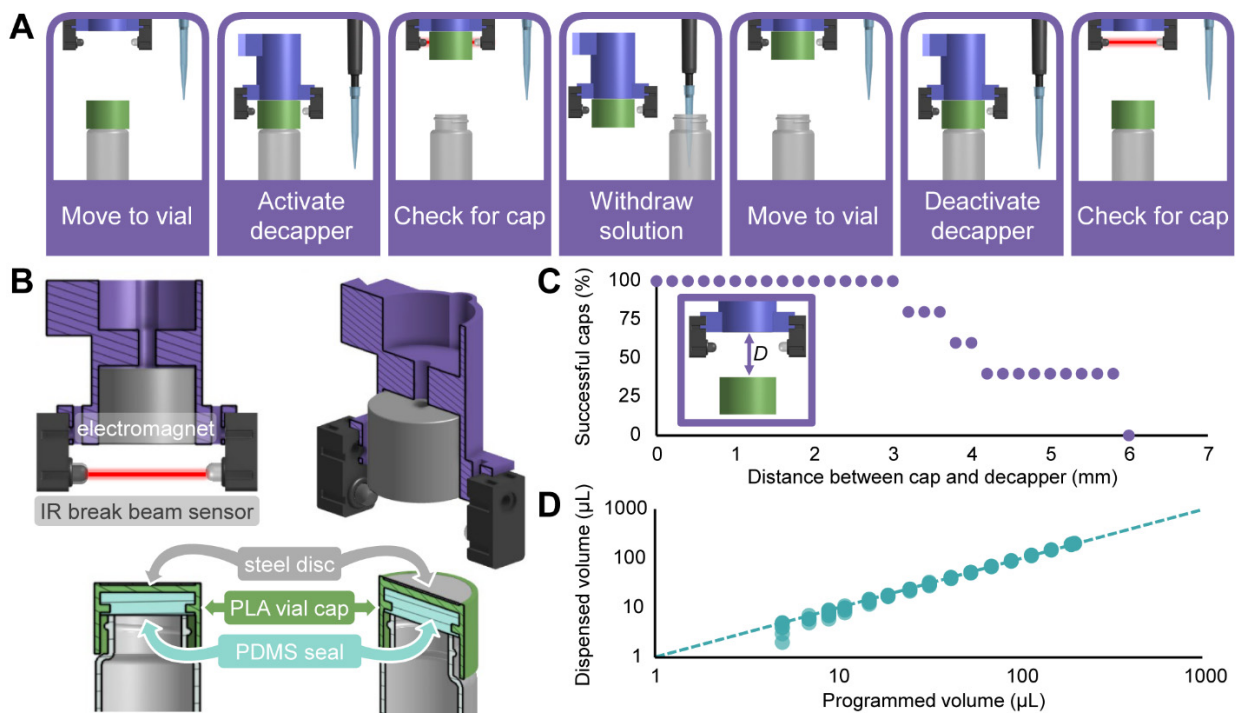

Figure 2: A. Process for using the electromagnetic capper/decapper for opening and closing a vial. B. Cross-section of a custom cap and the capper/decapper. C. Success rate of capper/decapper vs. programmed approach height showing tolerance to ~3 mm misalignment. D. Evaluation of fluid dispensing in terms of dispensed volume vs programmed volume to quantify accuracy/precision vs. dispensed volume (mass-based). Notable results include programmed 10 μL resulting in 9.7 ± 0.8 μL deposited and standard deviations ≲ 1 μL for programmed volumes ≥ 6 μL.

To validate the consistency and robustness of the capping/decapping process across independently prepared caps, we tested five caps using a series of programmed vertical approach heights, incremented in 0.2 mm steps. Each height was tested ten times for both decapping and subsequent recapping. As shown in Figure 2C, the system tolerated up to 3 mm of height misalignment without compromising performance, demonstrating robust and reproducible operation across multiple independently fabricated caps.

Following capper/decapper validation, we evaluated the performance of the Opentrons OT2 P300 single-channel pipette for liquid handling. While the manufacturer specifies an operating range of 20–300 μL, PANDA-film requires accurate dispensing down to 10 μL for the contact-angle assay as well as volumes up to 200 μL for well filling. To avoid additional complexity from integrating a second pipette, we prioritized using a single pipette for all liquid handling operations. Pipette accuracy and precision were assessed by dispensing defined volumes (2-200 µL) ten times each onto an analytical balance. Given the scale's readability (1 mg), repeatability (±1 mg), and linearity error (±2 mg), volumes below ~5 µL could not be reliably detected. At 10 µL, the pipette consistently dispensed 9.7 ± 0.8 µL, which aligns with the specifications in the Opentrons white paper[36] for 20 µL dispensing (±0.8 µL accuracy, ±0.5 µL precision) and within the acceptable range for goniometry. The dispensing accuracy was found to remain consistent with increasing volume dispensed, with standard deviations typically under 1 µL for volumes ≥ 6 µL (Figure 2D). These results confirm that the P300 pipette delivers sufficient accuracy and precision to support both low-volume and high-volume operations within PANDA-film. Compared to the syringe pump system used in the original PANDA, the pipette represents a significant performance improvement by virtue of being substantially faster (up to ~277 µL/s for the P300 vs 15 µL/s for the syringe pump), more precise (RMSE of 1.2 µL for the P300 pipette compared to 2.3 µL for the syringe pump), and capable of automated tip exchange.

## 3.2 Integration and Validation of the Contact Angle Assay

To enable automated characterization of surface wettability, we developed a top-down contact angle assay based on optical imaging of sessile water droplets deposited on thin films and substrates. Unlike conventional side-view goniometry, the developed approach infers droplet curvature by analyzing the reflection of paired LEDs projected onto the apex of the droplet (Figure 3A). The apparent separation between LED reflections decreases with increasing contact angle due to changes in droplet height and curvature, enabling a non-destructive, alignment-tolerant measurement. This configuration is particularly well-suited for high-throughput workflows using well plates, where side-view imaging is impractical.

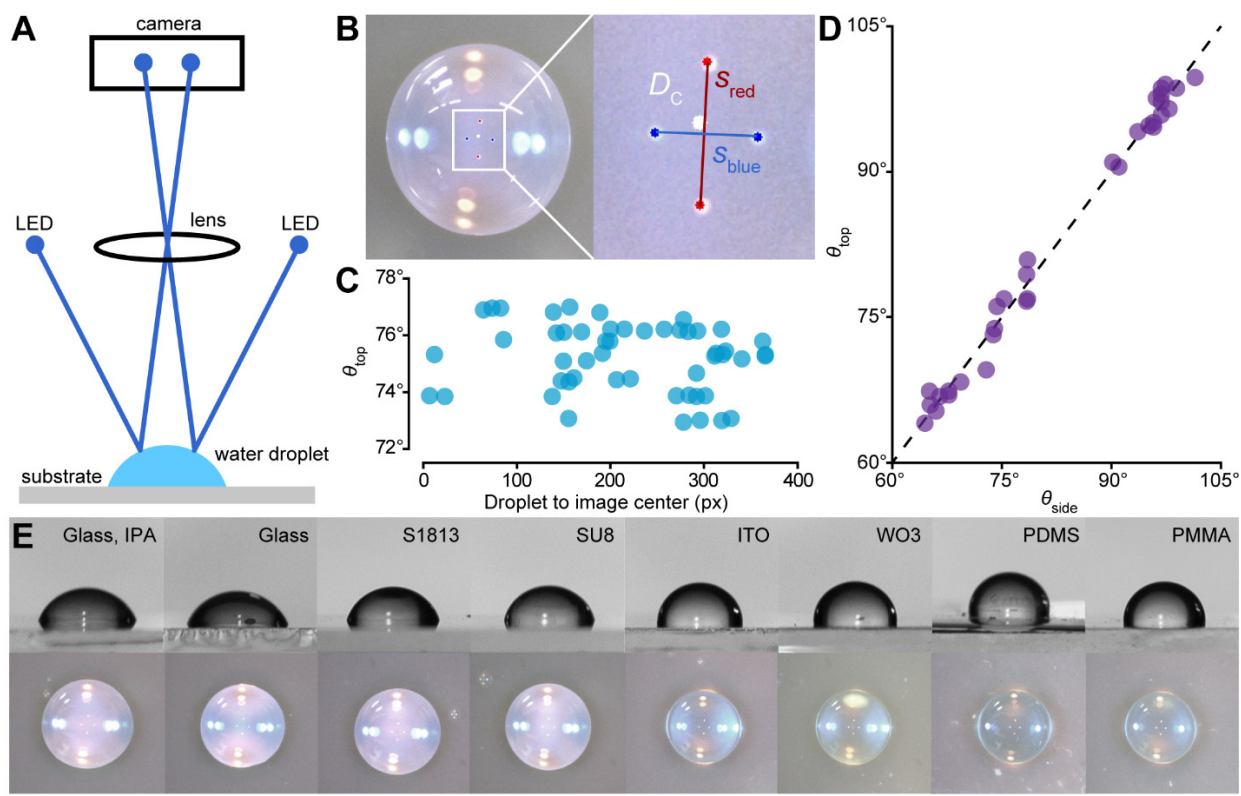

Figure 3: A. Top-down contact-angle concept using paired LED reflections on the droplet apex. B. Photograph showing the top of a droplet exhibiting reflections from two pairs of LEDs. Magnified region shows the image processing used to compute red LED separation $s_{red}$ and blue LED separation $s_{blue}$. C. Contact angle $\theta_{top}$ measured from the top down vs. lateral position of the droplet relative to the camera showing an insensitivity to precise camera position. The camera is moved 0.5 mm between each point. D. $\theta_{top}$ vs. side-video contact angle $\theta_{side}$ showing good agreement with parity. E. Calibration substrates showing top-down (bottom row, with reflections) and side-view (top row) images.

For model development and validation of the assay, 10 μL water droplets were dispensed onto cleaned substrates using the integrated pipette. Images were captured using a top-down camera while the droplet was illuminated by two pairs of LEDs (red and blue) (Figure 3B). To confirm droplet placement would not bias measurements, we conducted an experiment in which a droplet was imaged while the camera was moved laterally in 0.5 mm increments. No consistent trend in apparent contact angle was observed, confirming that the assay is robust to precise alignment between the camera and the droplet (Figure 3C). Validation experiments were then performed on eight different substrates, spanning a large range of water contact angles, namely SU-8, S1813, PDMS, bare glass, freshly rinsed glass, ITO, WO₃, and PMMA. For these experiments, droplets were placed and imaged sequentially in four positions on each substrate. Simultaneously, backlit side-view reference images were captured with a DSLR camera (Canon Rebel T5, 50 mm f/1.8 STM lens) (Figure S6). Top-down images were acquired over a 2 mm z-range with 0.25 mm increments to span the expected droplet height. Each z-slice was then evaluated by a Python script implementing a custom LED-aware focus score (combining Laplacian-variance sharpness, local contrast and compactness metrics, and red/blue LED geometric consistency) to select the optimal in-focus image. Reference contact angles from side-view images were measured using the Contact Angle plugin in ImageJ (National Institutes of Health). With this data in hand, we established an empirical calibration that converts top-down LED reflection spacings into a top-view contact angle $\theta_{top}$. For each droplet, we acquired a top-down image from which we measured the distances between the red and blue LED reflections in units of pixels, $s_{red}$ and $s_{blue}$, and a side-view goniometric measurement of $\theta_{side}$ simultaneously. Notably, this latter measurement is only done for calibration samples and not used during normal top-down operation. Using these paired measurements, we fit a quadratic regression model that maps the two reflection features to $\theta_{top}$,

$$\theta_{top} = As_{red}^2 + Bs_{red}s_{blue} + Cs_{blue}^2 + Ds_{red} + Es_{blue} + F, \tag{1}$$

where $A$, $B$, $C$, $D$, $E$, and $F$ are found using non-linear least squares fitting. After fitting the model using the full paired dataset, the model was saved and used to analyze top-down images to produce estimates of $\theta_{top}$ from ($s_{red}, s_{blue}$) enabling high-throughput analysis without the need for side-view imaging.

Across a diverse set of surface chemistries, the top-down method yielded contact angles within 5° of side-view measurements and performed reliably across both hydrophilic and hydrophobic surfaces, including conductive (ITO, WO₃) and dielectric (PDMS, SU-8, S1813, PMMA, glass) substrates. Agreement with side-view goniometry on the final calibration set ($n$ = 32) showed negligible bias (mean residual of 0.0°), RMSE = 1.3°, and MAE = 1.1°. Model performance, shown in Figure 3D, compares $\theta_{side}$ to $\theta_{top}$. Validating the top-down predictions using side-view measurements ensured that the resulting model was grounded in accurate, high-confidence data. Representative top-down and side-view images for each substrate are shown in Figure 3E.

This LED-based assay provides a rapid, compact, and reproducible alternative to side-profile imaging. Unlike conventional goniometers, it is compatible with well-plates and multiplexed formats common in SDLs, and it enables parallel spatial mapping of wettability across individual films, tasks that are challenging with traditional approaches.

**3.3 Demonstration of EPoN Film Deposition and Surface Characterization**

We sought to explore the ability of PANDA-film to study the water contact angle of electrodeposited polymer films. Conceptually, the process would entail depositing a polymer film using EPoN, rinsing the film using the automated pipetting system, and then measuring its water contact angle using the calibrated top-down contact-angle assay. Prior to performing experiments, however, it was necessary to ensure that contact angle measurements were performed on fully dry films. Thus, we first validated solvent evaporation on bare ITO wells rinsed using the same post-deposition protocol as in the deposition experiments (rinsing with DMF followed by rinsing with acetonitrile). Wells were imaged every minute until no residual solvent remained. Across three replicates, complete evaporation occurred within ~40 min, so all subsequent film measurements allowed 1 h for solvent evaporation.

Having validated conditions that would result in dry wells for characterization, we demonstrated integration of deposition and surface characterization in PANDA-film using a representative and EPoN-compatible polymer: phenol-modified PAMA. Polymer films were prepared in triplicate using the PANDA-film via EPoN by depositing the film at 1.8 V vs. AgQRE for 600 s. After deposition, rinsing, and drying, the water contact angle was measured by robotically dispensing a 10 μL droplet of water on the film and then imaging the droplet while illuminating it with the red and blue LEDs. Red/blue LED reflection spacings were extracted from top-down images and converted to $\theta_{top}$ via Eq. (1). Across the three polymer films, PANDA-film produced similar deposition profiles and transferred charge, confirming repeatability of EPoN on the PANDA-film system (Figure 4A). Thickness measurements likewise showed good agreement, as shown in Figure 4B. Average contact angles measured by PANDA-film were $103 \pm 0.6°$ illustrating similar surface wettability across replicates (Figure 4C). In addition to quantitative measurements, representative images are shown for each film. Figure 4D displays a top-down image captured with contact-angle illumination, highlighting the red and blue LED reflections used to infer $\theta_{top}$.

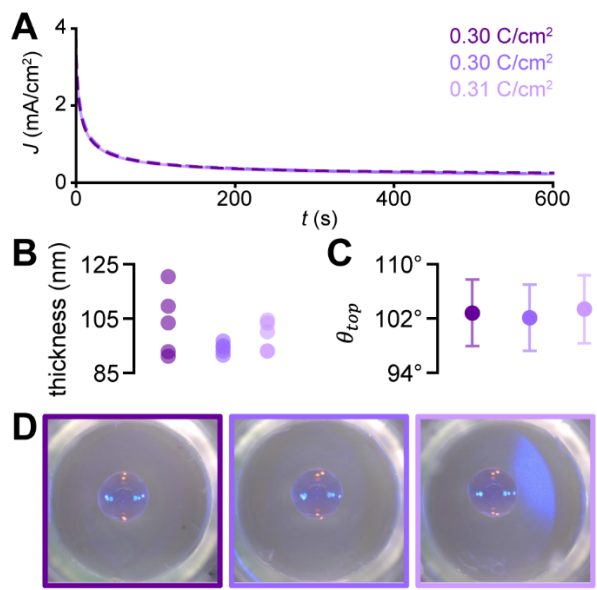

**Figure 4: A. Testing electrodeposition of polymer network (EPoN) using the PANDA-film. Charge density $J$ measured during deposition of phenol-modified PAMA vs. time $t$ with labels showing the corresponding integrated charge of each sample. B. Film thickness of each PAMA sample. C. $\theta_{top}$ measued for each PAMA sample. D Top-down reflection image of each PAMA sample.**

Together, these comparisons show that PANDA-film reproduces key electrochemical and thickness outcomes and delivers estimates of contact angle in an automated fashion with only a top-down image. By combining automated deposition, rinsing, drying, and top-down contact angle analysis, PANDA-film establishes a robust platform for investigating structure-property relationships in electrodeposited polymer thin films.

## 4. Conclusion

In this work, we present PANDA-film, a configuration of the PANDA platform designed for the automated synthesis and surface characterization of electrodeposited polymer films. The system integrates new hardware modules including a capping/decapping tool, a robotic pipette, and a top-down contact angle assay, collectively enabling high-throughput evaluation of film wettability in a well-plate format. Each module was validated independently, demonstrating reliable performance in capping consistency, liquid handling precision, and contact angle accuracy relative to conventional side-view goniometry. We further demonstrated PANDA-film by depositing films of a representative EPoN-compatible polymer. Comparison of deposition curves, charge passed, film thickness, and contact angle measurements showed consistency between replicates.

Looking forward, PANDA-film provides a foundation for systematic exploration of structure-property relationships in electrochemically synthesized films. Future studies could vary deposition potential along with concentration and electrolyte composition to map how processing governs thickness, morphology, and wettability. For example, if processing conditions impact surface roughness, intrinsic wetting (Wenzel-type behavior) may be amplified, lowering contact angle on hydrophilic films and raising contact angle on hydrophobic films.[30] Pairing PANDA-film's *in situ* top-down readout with interferometry for thickness, atomic force microscopy for roughness, and X-ray photoelectron spectroscopy for surface chemistry will enable quantitative links from deposition parameters to surface properties for both mechanistic insight and application-driven optimization.

**Supplementary Material**

Supplementary information includes nuclear magnetic resonance spectra of polymers, a photograph of the PANDA-film, a screen capture of the PANDA-film software interface, wiring diagrams of the PAW, and a photograph of side-view contact angle system used for reference.

**Acknowledgements**

The authors acknowledge support from the National Science Foundation (TIP-2229018 and CBET-2146597) and the Army Research Office (W911NF-25-20074 and W911NF-24-20197). The authors acknowledge support from the Photonics Center at Boston University.

**Conflicts of Interest**

The authors declare no conflicts of interest.

**Author Contributions**

Conceptualization: HQ, GAR, MP, NP, LP, JW, and KAB. Software: HQ and GAR. Investigation: HQ, GAR, ZZ, JY, YL, and ZY. Writing - Original Draft: HQ and KAB. Writing – Reviewing and Editing: All Authors.

**Data Availability**

All data is present in the manuscript. All software and 3D models are available at: https://github.com/BU-KABlab/PANDA-BEAR

**References**
(1) Abolhasani, M.; Kumacheva, E. The Rise of Self-Driving Labs in Chemical and Materials Sciences. *Nat. Synth.* **2023**, *2* (6), 483–492. https://doi.org/10.1038/s44160-022-00231-0.
(2) Stach, E.; DeCost, B.; Kusne, A. G.; Hattrick-Simpers, J.; Brown, K. A.; Reyes, K. G.; Schrier, J.; Billinge, S.; Buonassisi, T.; Foster, I.; Gomes, C. P.; Gregoire, J. M.; Mehta, A.; Montoya, J.; Olivetti, E.; Park, C.; Rotenberg, E.; Saikin, S. K.; Smullin, S.; Stanev, V.; Maruyama, B. Autonomous Experimentation Systems for Materials Development: A Community Perspective. *Matter* **2021**, *4* (9), 2702–2726. https://doi.org/10.1016/j.matt.2021.06.036.
(3) Snapp, K. L.; Verdier, B.; Gongora, A. E.; Silverman, S.; Adesiji, A. D.; Morgan, E. F.; Lawton, T. J.; Whiting, E.; Brown, K. A. Superlative Mechanical Energy Absorbing Efficiency Discovered through Self-Driving Lab-Human Partnership. *Nat. Commun.* **2024**, *15* (1), 4290. https://doi.org/10.1038/s41467-024-48534-4.

(4) Rooney, M. B.; MacLeod, B. P.; Oldford, R.; Thompson, Z. J.; White, K. L.; Tungjunyatham, J.; Stankiewicz, B. J.; Berlinguette, C. P. A Self-Driving Laboratory Designed to Accelerate the Discovery of Adhesive Materials. *Digit. Discov.* **2022**, *1* (4), 382–389. https://doi.org/10.1039/d2dd00029f.

(5) Dave, A.; Mitchell, J.; Burke, S.; Lin, H.; Whitacre, J.; Viswanathan, V. Autonomous Optimization of Non-Aqueous Li-Ion Battery Electrolytes via Robotic Experimentation and Machine Learning Coupling. *Nat. Commun.* **2022**, *13* (1), 5454. https://doi.org/10.1038/s41467-022-32938-1.

(6) Burger, B.; Maffettone, P. M.; Gusev, V. V.; Aitchison, C. M.; Bai, Y.; Wang, X.; Li, X.; Alston, B. M.; Li, B.; Clowes, R.; Rankin, N.; Harris, B.; Sprick, R. S.; Cooper, A. I. A Mobile Robotic Chemist. *Nature* **2020**, *583* (7815), 237–241. https://doi.org/10.1038/s41586-020-2442-2.

(7) Wang, C.; Kim, Y.-J.; Vriza, A.; Batra, R.; Baskaran, A.; Shan, N.; Li, N.; Darancet, P.; Ward, L.; Liu, Y.; Chan, M. K. Y.; Sankaranarayanan, S. K. R. S.; Fry, H. C.; Miller, C. S.; Chan, H.; Xu, J. Autonomous Platform for Solution Processing of Electronic Polymers. *Nat. Commun.* **2025**, *16* (1). https://doi.org/10.1038/s41467-024-55655-3.

(8) Du, X.; Lüer, L.; Heumueller, T.; Wagner, J.; Berger, C.; Osterrieder, T.; Wortmann, J.; Langner, S.; Vongsaysy, U.; Bertrand, M.; Li, N.; Stubhan, T.; Hauch, J.; Brabec, C. J. Elucidating the Full Potential of OPV Materials Utilizing a High-Throughput Robot-Based Platform and Machine Learning. *Joule* **2021**, *5* (2), 495–506. https://doi.org/10.1016/j.joule.2020.12.013.

(9) MacLeod, B. P.; Parlane, F. G. L.; Morrissey, T. D.; Häse, F.; Roch, L. M.; Dettelbach, K. E.; Moreira, R.; Yunker, L. P. E.; Rooney, M. B.; Deeth, J. R.; Lai, V.; Ng, G. J.; Situ, H.; Zhang, R. H.; Elliott, M. S.; Haley, T. H.; Dvorak, D. J.; Aspuru-Guzik, A.; Hein, J. E.; Berlinguette, C. P. Self-Driving Laboratory for Accelerated Discovery of Thin-Film Materials. *Sci. Adv.* **2020**, *6* (20). https://doi.org/10.1126/sciadv.aaz8867.

(10) Vasquez, J.; Twigg-Smith, H.; Tran O'Leary, J.; Peek, N. Jubilee: An Extensible Machine for Multi-Tool Fabrication. In *Proceedings of the 2020 CHI Conference on Human Factors in Computing Systems*; ACM: Honolulu HI USA, 2020; pp 1–13. https://doi.org/10.1145/3313831.3376425.

(11) Epps, R. W.; Volk, A. A.; Ibrahim, M. Y. S.; Abolhasani, M. Universal Self-Driving Laboratory for Accelerated Discovery of Materials and Molecules. *Chem* **2021**, *7* (10), 2541–2545. https://doi.org/10.1016/j.chempr.2021.09.004.

(12) Zheng, Z.; Resing, A. B.; Wang, W.; Werner, J. G. Cathodic Electrodeposition of Polymer Networks as Ultrathin Films on 3-D Micro-Architected Electrodes. *RSC Appl. Polym.* **2024**, *2* (6), 1139–1146. https://doi.org/10.1039/D4LP00180J.

(13) Palma-Cando, A.; Rendón-Enríquez, I.; Tausch, M.; Scherf, U. Thin Functional Polymer Films by Electropolymerization. *Nanomaterials* **2019**, *9* (8), 1125. https://doi.org/10.3390/nano9081125.

(14) Wang, W.; Resing, A. B.; Brown, K. A.; Werner, J. G. Electrodeposition of Polymer Networks as Conformal and Uniform Ultrathin Coatings. *Adv. Mater.* **2024**, *36* (48), 2409826. https://doi.org/10.1002/adma.202409826.

(15) Quinn, H.; Robben, G.; Zheng, Z.; Gardner, A.; Werner, J. G.; Brown, K. A. PANDA: A Self-Driving Lab for Studying Electrodeposited Polymer Films. *Mater. Horiz.* **2024**, 10.1039.D4MH00797B. https://doi.org/10.1039/D4MH00797B.

(16) Pittaway, P. M.; Knox, S. T.; Cayre, O. J.; Kapur, N.; Golden, L.; Drillieres, S.; Warren, N. J. Self-Driving Laboratory for Emulsion Polymerization. *Chem. Eng. J.* **2025**, *507*, 160700. https://doi.org/10.1016/j.cej.2025.160700.

(17) Taherimakhsousi, N.; MacLeod, B. P.; Parlane, F. G. L.; Morrissey, T. D.; Booker, E. P.; Dettelbach, K. E.; Berlinguette, C. P. Quantifying Defects in Thin Films Using Machine Vision. *Npj Comput. Mater.* **2020**, *6* (1), 111. https://doi.org/10.1038/s41524-020-00380-w.

(18) Meredith, J. C.; Smith, A. P.; Karim, A.; Amis, E. J. Combinatorial Materials Science for Polymer Thin-Film Dewetting. *Macromolecules* **2000**, *33* (26), 9747–9756. https://doi.org/10.1021/ma001298g.

(19) Subedi, D. P. Contact Angle Measurement for The Surface Characterization of Solids. *Himal. Phys.* **2011**, *2*, 1–4. https://doi.org/10.3126/hj.v2i2.5201.
(20) Żenkiewicz, M. Comparative Study on the Surface Free Energy of a Solid Calculated by Different Methods. *Polym. Test.* **2007**, *26* (1), 14–19. https://doi.org/10.1016/j.polymertesting.2006.08.005.
(21) Hubbe, M. A.; Gardner, D. J.; Shen, W. Contact Angles and Wettability of Cellulosic Surfaces: A Review of Proposed Mechanisms and Test Strategies. *BioResources* **2015**, *10* (4), 8657–8749. https://doi.org/10.15376/biores.10.4.8657-8749.
(22) Chau, T. T. A Review of Techniques for Measurement of Contact Angles and Their Applicability on Mineral Surfaces. *Miner. Eng.* **2009**, *22* (3), 213–219. https://doi.org/10.1016/j.mineng.2008.07.009.
(23) Hegemann, D.; Brunner, H.; Oehr, C. Plasma Treatment of Polymers for Surface and Adhesion Improvement. *Nucl. Instrum. Methods Phys. Res. Sect. B Beam Interact. Mater. At.* **2003**, *208*, 281–286. https://doi.org/10.1016/S0168-583X(03)00644-X.
(24) Nemani, S. K.; Annavarapu, R. K.; Mohammadian, B.; Raiyan, A.; Heil, J.; Haque, Md. A.; Abdelaal, A.; Sojoudi, H. Surface Modification of Polymers: Methods and Applications. *Adv. Mater. Interfaces* **2018**, *5* (24), 1801247. https://doi.org/10.1002/admi.201801247.
(25) Nathanael, A. J.; Oh, T. H. Biopolymer Coatings for Biomedical Applications. *Polymers* **2020**, *12* (12), 3061. https://doi.org/10.3390/polym12123061.
(26) Giljean, S.; Bigerelle, M.; Anselme, K.; Haidara, H. New Insights on Contact Angle/Roughness Dependence on High Surface Energy Materials. *Appl. Surf. Sci.* **2011**, *257* (22), 9631–9638. https://doi.org/10.1016/j.apsusc.2011.06.088.
(27) Drummond, C. J.; Chan, D. Y. C. Van Der Waals Interaction, Surface Free Energies, and Contact Angles: Dispersive Polymers and Liquids. *Langmuir* **1997**, *13* (14), 3890–3895. https://doi.org/10.1021/la962131c.
(28) Huhtamäki, T.; Tian, X.; Korhonen, J. T.; Ras, R. H. A. Surface-Wetting Characterization Using Contact-Angle Measurements. *Nat. Protoc.* **2018**, *13* (7), 1521–1538. https://doi.org/10.1038/s41596-018-0003-z.
(29) Guilizzoni, M. Drop Shape Visualization and Contact Angle Measurement on Curved Surfaces. *J. Colloid Interface Sci.* **2011**, *364* (1), 230–236. https://doi.org/10.1016/j.jcis.2011.08.019.
(30) Meiron, T. S.; Marmur, A.; Saguy, I. S. Contact Angle Measurement on Rough Surfaces. *J. Colloid Interface Sci.* **2004**, *274* (2), 637–644. https://doi.org/10.1016/j.jcis.2004.02.036.
(31) Bartell, F. E.; Shepard, J. W. The Effect of Surface Roughness on Apparent Contact Angles and on Contact Angle Hysteresis. I. The System Paraffin–Water–Air. *J. Phys. Chem.* **1953**, *57* (2), 211–215. https://doi.org/10.1021/j150503a017.
(32) Dutra, G.; Canning, J.; Padden, W.; Martelli, C.; Dligatch, S. Large Area Optical Mapping of Surface Contact Angle. *Opt. Express* **2017**, *25* (18), 21127. https://doi.org/10.1364/OE.25.021127.
(33) Janeczko, C.; Martelli, C.; Canning, J.; Dutra, G. Assessment of Orchid Surfaces Using Top-Down Contact Angle Mapping. *IEEE Access* **2019**, *7*, 31364–31375. https://doi.org/10.1109/ACCESS.2019.2902730.
(34) Wang, W.; Li, Y.; Resing, A. B.; Yan, J.; Zheng, Z.; Werner, J. G. Electrodeposition of Reactive Polymer Networks for Conformal Ultrathin Coatings Amenable to Post-Deposition Functionalization. *J. Mater. Chem. A* **2025**, *13* (35), 29050–29059. https://doi.org/10.1039/D5TA03811A.
(35) Wang, W.; Zheng, Z.; Resing, A. B.; Brown, K. A.; Werner, J. G. Conformal Electrodeposition of Ultrathin Polymeric Films with Tunable Properties from Dual-Functional Monomers. *Mol. Syst. Des. Eng.* **2023**, *8* (5), 624–631. https://doi.org/10.1039/D2ME00246A.
(36) Opentrons. Opentrons Single Pipette GEN2, 2019. https://s3.amazonaws.com/opentrons-landing-img/pipettes/OT-2-Pipette-White-Paper.pdf.